\documentclass{article}
\usepackage{spconf,amsmath,graphicx}
\usepackage{xcolor,booktabs}
\usepackage{hyperref}
\title{Zero-Shot Audio Captioning via Audibility Guidance}
\name{Tal Shaharabany, Ariel Shaulov and Lior Wolf}
\address{School of Computer Science, Tel Aviv University}
\begin{document}
%\ninept
%
\maketitle
\begin{abstract}
The task of audio captioning is similar in essence to tasks such as image and video captioning. However, it has received much less attention. We propose three desiderata for captioning audio -- (i) fluency of the generated text, (ii) faithfulness of the generated text to the input audio, and the somewhat related (iii) audibility, which is the quality of being able to be perceived based only on audio. Our method is a zero-shot method, i.e., we do not learn to perform captioning. Instead, captioning occurs as an inference process that involves three networks that correspond to the three desired qualities: (i) A Large Language Model, in our case, for reasons of convenience, GPT-2, (ii) A model that provides a matching score between an audio file and a text, for which we use a multimodal matching network called ImageBind, and (iii) A text classifier, trained using a dataset we collected automatically by instructing GPT-4 with prompts designed to direct the generation of both audible and inaudible sentences. %, taking into account factors such as word coherence and grammatical correctness. 
We present our results on the AudioCap dataset, demonstrating that audibility guidance significantly enhances performance compared to the baseline, which lacks this objective.
\end{abstract}
\section{Introduction}
\label{sec:intro}

Audio captioning is the task of automatically generating a natural language description of an audio clip. This is a challenging task since audio can be ambiguous and difficult to interpret. However, audio captioning has a number of potential applications, such as helping the hearing-impaired understand the world around them and creating indexable and accessible multimedia content.

We propose a novel zero-shot audio captioning method that addresses the challenges of fluency, faithfulness, and audibility. Our method employs transfer learning from three neural networks: (i) A large language model (LM) that generates fluent text and incorporates world knowledge, (ii) A multimodal matching network that scores the similarity between an audio clip and a potential text description, and (iii) A text classifier that predicts whether a caption is audible.

While components that are similar or analogous to the first two networks are commonplace in image captioning, the last component is entirely novel. It is used to make sure that the generated text describes something that can be inferred from the audio itself. Otherwise, the language model can generate a sentence that is fluent and describes the content of the audio but adds elements that cannot be inferred from the audio. For example, the generated text could refer to barking by a black dog.

\begin{figure}[t]
    \centering
    \includegraphics[width=1.05\linewidth]{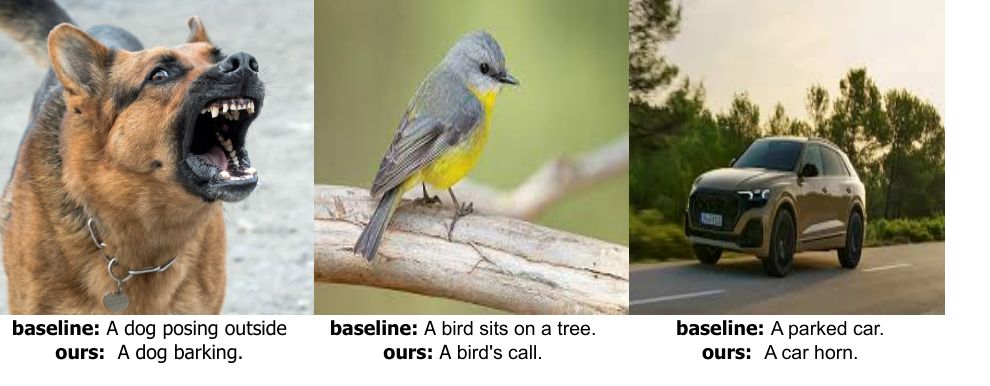}
    \vspace{-8mm}
    \caption{In the context of zero-shot audio captioning, we present illustrative instances of caption generation produced by our pipeline. It is important to emphasize that our algorithm operates in a fully blind manner, relying solely on audio inputs (the video frames are for illustration). Each sample is accompanied by two captions: our method generates an audible description, while the baseline, which lacks our audibility guidance, produces an inaudible description.}
    \label{fig:teaser}
    \vspace{-4mm}
\end{figure}

This pitfall is, as we show, very common, and stems from two factors: (i) the LM is trained on texts that are predominantly generated by visually-dominated creatures to fit the literary standard of such creatures~\cite{hutmacher2019there}, and (ii) multimodal match-scoring models are trained via a contrastive learning scheme~\cite{oord2018representation}, which does not directly penalize for captions that contain superfluous details. 

Fig.~\ref{fig:teaser} depicts sample audios (represented by corresponding video frame, which does not play a part in the audio caption), along with their corresponding captions, showcasing two modes of guidance: our full method, which includes the audibility objective and a baseline method (ablation) that lacks this term. As can be seen, without the audibility term, captioning generates sentences that describe elements that are hard or impossible to hear. With the introduction of the audibility term, the output becomes audio-appropriate. %These instances collectively emphasize the pivotal role of audible guidance in the fully blind caption generation process facilitated by our algorithm.

Our contributions to the field of audio captioning are:
(i) We introduce an algorithm that represents the first known endeavor in the domain of zero-shot audio captioning.
(ii) We propose a novel approach to enhancing the audibility of generated captions through the incorporation of classifier guidance. 
(iii) We introduce a carefully curated dataset specifically generated by ChatGPT that contains audible and inaudible caption samples. %for evaluating the quality of audio captions with respect to their audibility.

We evaluate our method on the AudioCap~\cite{kim2019audiocaps} dataset, showcasing the effectiveness of audibility guidance in improving performance compared to the baseline, in both supervised metrics, which compare to the ground truth caption and indicate the quality of audibility, respectively.
% {\color{red}TO BE UPDATED We evaluate our method on a number of benchmark datasets, and show that it outperforms state-of-the-art methods in terms of all three desiderata.
% }
\section{Related Work}

Captioning, in various forms, has been extensively studied for images and videos. However, the same cannot be said for audio content, especially in the realm of zero-shot learning, and as far as we can ascertain, our work is the first to tackle zero-shot audio captioning. %To the best of our knowledge, our work presents the pioneering efforts in the domain of zero-shot audio captioning. This section delves into the related literature spanning audio captioning, zero-shot captioning, classifier guidance, and classifier-free guidance.

\noindent\textbf{Audio Captioning\quad} Historically, audio captioning research has predominantly adopted the fully-supervised encoder-decoder architectures~\cite{mei2021audio, liu2022leveraging}.
Such methodologies extract acoustic characteristics using an audio encoder and, subsequently, deploy a language decoder to generate captions. Kadlvcik et al. proposed a noteworthy approach that harnessed the capabilities of Whisper's~\cite{kadlvcik2023whisper} transformer for audio preprocessing~\cite{radford2023robust}. Concurrently, Liu et al. demonstrated the efficacy of integrating pre-trained BERT~\cite{devlin2018bert} for decoder initialization, complemented by the adoption of pre-trained Audio Neural Networks (PANNs) as encoders. A recent trend is the application of audio captioning within the framework of multi-task methodologies~\cite{chen2023valor, deshmukh2023pengi}.

\noindent\textbf{Zero-Shot Captioning\quad} Aligning closely with our contributions, the arena of zero-shot captioning seeks to conjure meaningful and syntactically coherent descriptions for various input modalities, be it images, videos, or audio.
Su et al.~\cite{su2022language} and Tewel et al.~\cite{tewel2022zerocap} performed zero-shot image captioning, capitalizing on the prowess of large pre-trained entities, such as CLIP~\cite{radford2021learning} and GPT~\cite{radford2019language}. In a similar vein, Bracha et al. extended the approach to the domain of zero-shot Referring Expressions Generation, focusing on crafting unambiguous textual elucidations of visual entities~\cite{bracha2023disclip}. The contemporaneous work by Tewel et al. unveiled a strategy for zero-shot video captioning, enriched by the incorporation of pseudo-tokens to dynamically update context~\cite{tewel2022zero}.

\begin{table*}[h]
    \centering
    \resizebox{0.74\textwidth}{!}{ % Add this line to resize the table to fit the text width
        \begin{tabular}{cc}
            \hline
            Audible & Not Audible \\
            \hline
            The barking of a dog in excitement. & Magnets attract metals. \\
            Ringing phone awaits an answer. & Ice covers the lake in winter. \\
            The buzz of a drone flying overhead. & Icebergs float on water. \\
            Birds chirping in early morning. & An umbrella stands closed by the door. \\
            Jingling coins are counted or played with. & A statue in a park. \\
            The swishing sound of a washing machine. & Rusting car sits in the yard. \\
            Whips cracked in the rodeo. & Resolved issue is fixed. \\
            The meow of a cat. & Soccer balls are stored in a mesh bag. \\
            The school bell rings, signaling the end of class. & Skimmed milk has less fat. \\
            Raindrops tapping on rooftops. & A pair of hiking boots rests next to a backpack. \\
            \hline
        \end{tabular}
    }
    \vspace{-2mm}
    \caption{Examples from the GPT-4 generated dataset comprising both audible and inaudible sentences. We prompted ChatGPT with the task of creating examples for a classifier that distinguishes between these two categories.} 
    % The prompts also instructed ChatGPT to consider factors such as word coherence, grammatical correctness, context, and the likelihood of the sentence representing a meaningful auditory scenario. 
    \label{tab:my_table}
    \vspace{-5mm}
\end{table*}

% Our work is related to the literature on zero-shot captioning, where the goal is to describe an input (image, video, audio, etc.) with meaningful and syntactically correct sentences. This topic has been explored from several directions in the past. \cite{su2022language,tewel2022zerocap} suggests a method for zero-shot image captioning in which they are acquiring two large pre-trained models: CLIP and GPT. A similar method has been suggested by \cite{bracha2023disclip} for zero-shot Referring Expressions Generation, which aims to produce textual descriptions that unambiguously identify specific objects within a visual scene. In recent work \cite{tewel2022zero} suggests a method for zero-shot video captioning which involves pseudo-tokens to update the context. \\

\noindent\textbf{Classifier guidance\quad} Classifier guidance was introduced as a way to trade off mode coverage
and sample fidelity during the inference phase in pretrained Diffusion models~\cite{dhariwal2021diffusion}. In computer vision, Shi et al propose a novel paradigm of using the diffusion model and classifier guidance in the latent semantic space for compositional visual tasks~\cite{shi2023exploring}, and Epstein et al. propose Self-guidance that operates similarly to standard classifier guidance, but uses signals present in the pre-trained model itself~\cite{epstein2023diffusion}. Our work aims at using classifier guidance to guide models toward more audible captions.

\section{Method}

% {\color{red}JUST AN EXAMPLE TO GIVE YOU SOME GENERAL STRUCTURE AND INSPIRATION. DO NOT COPY TEXT AS IS BUT YOU CAN USE THE RELEVANT EQUATIONS}

% Our goal is to create a sentence $S=\{t_1,\ldots, t_M\}$ of length $M$ that describes a set of video frames $\mathcal{F}=\{F_1,\ldots, F_N\}$, where $N$ is the number of frames. When $N=1$, the problem corresponds to traditional image captioning.

The task of captioning can be mathematically framed as a sequence generation problem, wherein we seek to infer the conditional probability of the i-th word, denoted as $x_i$, in the sentence. In essence, our goal is to optimize the probability distribution p($x_i$|[$x_t$]$_{t<i}$, $\mathcal{A}$), where $x_t$ represents preceding words in the sentence, and $\mathcal{A}$ signifies the input - in our case, an audio clip.

We introduce a zero-shot approach specifically designed for the audio domain, employing a large-scale text-audio alignment model in conjunction with audibility promoting classifier-based guidance that together steer a large-scale language model.

\noindent\textbf{Main Pipeline\quad}
The proposed framework is composed of three fundamental networks: (i) a large language model (LM), specifically GPT-2~\cite{radford2019language}, (ii) a multi-modal model called ImageBind~\cite{girdhar2023imagebind}, which is trained to establish alignment between the representation of an audio segment and a textual counterpart, and (iii) a binary audibility classifier designed to provide guidance in the captioning process.

The GPT-2 LM is constructed using  $L$ layers of Transformers, with each layer incorporating key and value embeddings to capture token interactions. In the transformer block, there are three embedding functions: $K$, $Q$, and $V$. These functions enable token interactions, with $Q$ determining the queries, $K$ the keys, and the attention mechanism pooling values $V$ based on the similarity between queries and keys, resulting in weighted average value vectors.

These key and value representations are stored in a context cache to retain a record of past embeddings. Consequently, the (unconditioned) sentence generation process is expressed as $x_{i+1} = \operatorname{LM}\left( x_i, [(K_j^l, V_j^l)]_{j<i,1\leq l\leq L} \right)$, where $x_i$ denotes the i-th word in the generated sentence, and $K_j^l$ and $V_j^l$ pertain to the context transformer's key and value for the j-th token across L layers. 

\noindent\textbf{Multi-Modal Network\quad}
Multi-modal networks typically comprise two distinct encoders, namely, a textual encoder denoted as $F_t$ and an audio encoder represented as $F_a$. In our approach, based on ImageBind~\cite{girdhar2023imagebind}, the primary objective of the multi-modal network denoted as $F$ is to offer audio-centric guidance to the Language Model (LM). This guidance is achieved by minimizing the following term:

\begin{equation}
\label{eq:one}
L_{mm}=  -\frac{F_t(LM(x_i;C_i)) \cdot F_a(\mathcal{A})}{\left\vert F_t(LM(x_i;C_i)) \right\vert \cdot \left\vert F_a(\mathcal{A}) \right\vert}\,, 
\end{equation}
where $C_i = [(K_j^l, V_j^l)]_{j<i,1\leq l\leq L}$ is the context cache defined above, over which we optimize $L_{mm}$ (and all our loss terms below). This form of optimizing the cache follows contributions such as~\cite{Dathathri2020Plug}. Our aim is to manipulate the outputs of the LM by considering their interaction with audio encoding, thereby enhancing the match between the generated text and the input audio.

% whare $\mathcal{A}$ is the 
% by iteratively assigning scores to the LM's output sentences. The scorfes are determined according to the cosine similarity between the generated text features and the query audio features.   % One could posit that the multi-modal network imparts the Language Model (LM) with the capacity to render its output audible.

\noindent\textbf{Audibility classifier\quad}
ImageBind~\cite{girdhar2023imagebind} matched audio, video, and text. The video dominates the training process, and the network optimization primarily revolves around aligning the visual video frames with the original video soundtrack. Therefore, when we employed ImageBind networks to guide GPT-2, we observed that the generated descriptions tend to prioritize visual elements at the expense of audibility, which is contrary to our objective of achieving audio captioning. Such captions cannot describe content, such as color that cannot be inferred from audio.

To address this challenge, we took the approach of generating two distinct sets of sentences through ChatGPT: one set representing audible captions and the other non-audible captions. This newly generated dataset serves as the foundation for training a classifier ($h_a$), which is specifically designed to distinguish between audible and non-audible captions.

In Table~\ref{tab:my_table}, we provide examples of sentences from the dataset used for training the classifier to distinguish between audible and non-audible descriptions. These sentences were generated using the following prompt to ChatGPT: ``I want to train a classifier that distinguishes between an audio description that is more audible and not audible - two classes. generate examples for each category. You should consider factors such as the coherence of words, grammatical correctness, context, and the likelihood that the sentence represents a meaningful auditory scenario.''.%, which requested examples representing two distinct classes: audible and not audible descriptions. The dataset incorporates considerations such as word coherence, grammatical correctness, context, and the likelihood that the sentence represents a meaningful auditory scenario.

The trained classifier $h_a$ offers guidance to the Language Model (LM) in a manner that aligns with our audibility objectives. This guidance is attained through optimizing the following term over the context cache $C_i$:

\begin{equation}
\label{eq:two}
L_{aud} = -\log\left(h_a(LM(x_i;C_i)[1]\right)
\end{equation}
Where $[1]$ indexes the classifier's output for the pseudo-probability for the audibility label (the positive label).

For $h_a$ we used DistilBERT architecture~\cite{sanh2019distilbert}, which is effective for the dataset we have collected (15,385 samples).

\noindent\textbf{Loss Function\quad}
Our methodology employs GPT-2 to deduce the subsequent word in a sentence, starting from an initial prompt ``Audio of a''.
To seamlessly integrate audio-related knowledge into the auto-regression process, we introduce a calibrated multi-modal loss term denoted as $L_{mm}$ (Eq.~\ref{eq:one}), which serves to incentivize the model to produce sentences that effectively describe a given audio segment. 

In order to compel the Language Model (LM) to produce sentences that are audible, we introduce an additional term referred to as $L_{aud}$ of Eq.~\ref{eq:two} into the primary loss function.

Furthermore, an additional regularization term denoted as $L_{CE}$ (Eq.~\ref{eq:three}) is added (as is often done) to ensure that the distribution of the next token remains consistent with that of the original language model.

\begin{equation}
\label{eq:three}
L_{CE} = CE(LM(x_i;C_i),LM(x_i;C^o_i))\,,
\end{equation}
where $i$ is the index of the currently generated token, CE is the cross entropy loss, and $C^o_i$ is a context cache of the relevant keys, queries, and values as they are computed based on the embedding and $K$ $Q$ and $V$ projections of the GPT-2 model (without the inference time optimization over $C_i$).

To conclude, the loss function of the optimization process can be represented as:
\begin{equation}
\label{eq:four}
   \mathcal{L} = \lambda_0\cdot\mathcal{L}_{CE} + \lambda_1\cdot\mathcal{L}_{mm} + \lambda_2\cdot\mathcal{L}_{aud},
\end{equation}

As default parameters, we set $\lambda_0$ to be 0.2, $\lambda_1$ to be 1 and $\lambda_2$ to be 0.015.
This optimization process is executed iteratively during auto-regression, with each token being addressed in sequence. At every generation step, we optimize the current context cache $C_i$ using gradient descent (see implementation details below), generate the next token, and continue to the next iteration.

\section{Experiments}

\begin{figure}[t]
    \centering
    \includegraphics[width=1.00\linewidth]{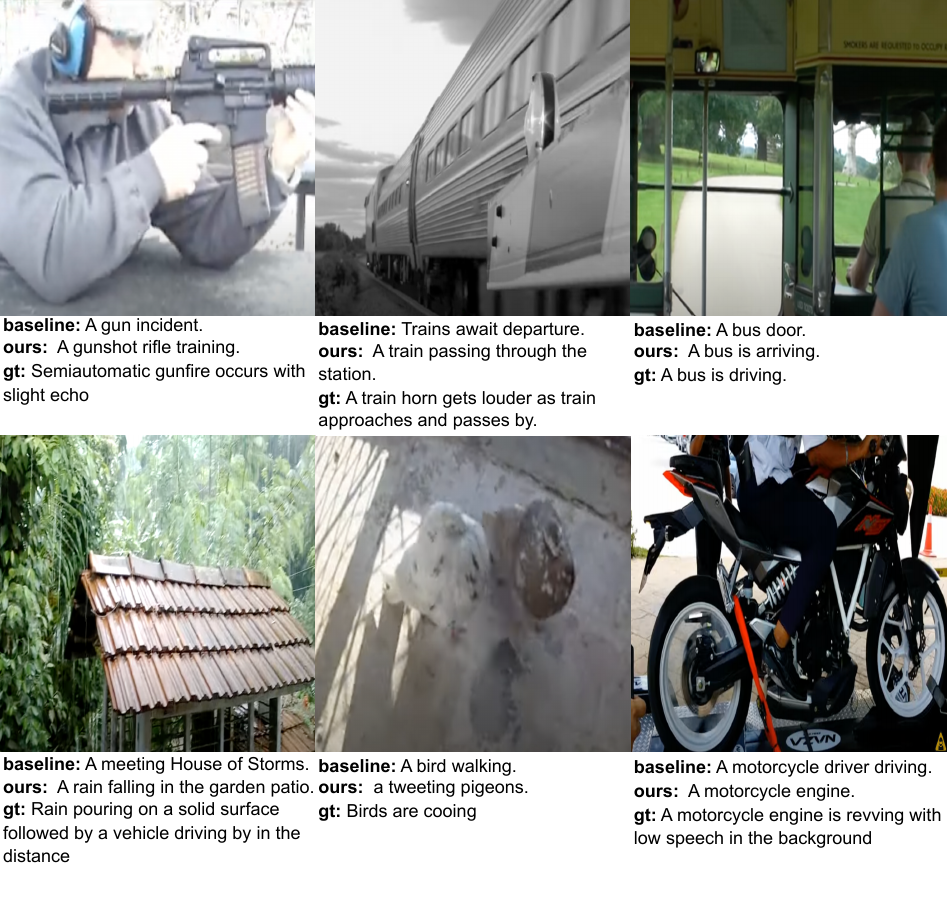}
    \vspace{-7mm}
    \caption{Examples of generated captions without (baseline) and with (ours) the audibility guidance, compared to the GT.}
    \label{fig:viz}
    \vspace{-5mm}
\end{figure}

For evaluating our solution, we utilize the test set (975 audio files) of the AudioCaps~\cite{kim2019audiocaps} dataset, which is specifically curated to assess audio captioning methodologies. AudioCaps~\cite{kim2019audiocaps} contains diverse audio segments, each paired with a reference textual description, which serves as the ground truth during evaluation. We train our audible classifier using our self-collected dataset, which contains two categories: audible, and non-audible.
% {Ariel ADD more details here please about the dataset}.

\noindent\textbf{Implementation details\quad}
Our use of the GPT-2 Large Language Model (LM) involves several key configuration and training aspects. We employ the GPT-2 model, which consists of 24 layers of Transformers. 
The pipeline runs on a single Titan X GPU and generates using a single beam search and 512 candidate tokens within a mere 8 seconds per token. We set the target sequence length to be 8.
% The inference time scales proportionally with the number of candidates employed.

The classifier $h_a$ is based on the DistilBert~\cite{sanh2019distilbert} network. The optimizer is AdamW with a batch size of 64 and a learning rate of 0.0003 for 40 epochs. The scheduler decreases the learning rate by a factor of 10 every 10 epochs. The audible dataset consists of 15,385 generated captions from ChatGPT: 8,383 for the not-audible category and 7,002 for the audible category. The audible category contains as little visual information as possible, as opposed to the non-audible category which contains as much visual information as possible.

\noindent\textbf{Evaluation} We evaluate our pipeline using standard fully supervised metrics for captioning assessments. These metrics include BLEU~\cite{papineni2002bleu}, METEOR~\cite{banerjee2005meteor}, ROUGE-L~\cite{lin2004rouge}, and CIDEr~\cite{vedantam2015cider}.
In addition to the conventional captioning metrics, we introduce a metric focused on audibility assessment. This metric leverages the predictions of the audibility classifier denoted as $h_a$. Specifically, we calculate the probability assigned by $h_a$ to each generated caption, representing the likelihood of the caption being audibly descriptive.

\begin{table}[h]
\centering
\resizebox{0.36\textwidth}{!}{%
\begin{tabular}{lccccc}
\hline
\textbf{Metric} & \textbf{B} & \textbf{M} & \textbf{R} & \textbf{C} & \textbf{A} \\
\hline
$\lambda_2=0$ & 9.3\% & 7.9\% & 7.4\% & 7.9\% & 60.9\% \\
$\lambda_2=0.015$ & \textbf{9.8\%} & 8.6\% & \textbf{8.2}\% & \textbf{9.2}\% & 62.8\% \\
$\lambda_2=0.03$ & 9.4\% & \textbf{8.8\%} & 7.8\% & 8.9\% & 69.8\% \\
$\lambda_2=0.06$ & 9.5\% & 8.3\% & 8.1\% & 8.5\% & 85.6\% \\
$\lambda_2=0.09$ & 9.5\% & 8.2\% & 7.8\% & 8.5\% & 91.9\% \\
$\lambda_2=0.12$ & 9.4\% & 8.3\% & 7.8\% & 8.4\% & 94.2\% \\
$\lambda_2=0.15$ & 9.4\% & 8.4\% & 7.7\% & 8.3\% & 95.8\% \\
$\lambda_2=0.2$ & 9.2\% & 8.1\% & 7.6\% & 7.4\% & 97.1\% \\
$\lambda_2=0.25$ & 9.5\% & 8.2\% & 7.9\% & 7.5\% & 97.9\% \\
$\lambda_2=0.30$ & 9.2\% & 8.0\% & 7.6\% & 7.3\% & 98.3\% \\
$\lambda_2=0.35$ & 9.4\% & 8.1\% & 7.8\% & 7.2\% & 98.8\% \\
$\lambda_2=0.40$ & 9.0\% & 7.8\% & 7.4\% & 6.4\% & 98.7\% \\
$\lambda_2=0.45$ & 9.0\% & 7.7\% & 7.5\% & 7.0\% & 98.8\% \\
$\lambda_2=0.50$ & 9.2\% & 7.7\% & 7.4\% & 7.2\% & \textbf{99.0\%} \\
\hline
\end{tabular}%
}
\caption{Evaluation results on the AudioCaps dataset for varying values of the weight parameter $\lambda_2$. The table displays several performance metrics, including BLEU (B), METEOR (M), ROUGE-L (R), and CIDEr (C), alongside a novel metric focused on audibility (A).}
\label{tab:results}
\vspace{-5mm}
\end{table}

\noindent\textbf{Results\quad}
The evaluation results from our pipeline on the AudioCaps dataset are presented in Tab.~\ref{tab:results} for different values of $\lambda_2$, associated with the audibility loss term $L_{aud}$. As can be seen, for a wide range of parameter values $[0.015, 0.15]$, all four supervised metrics improve over the $\lambda_2=0$ baseline, which lacks this term. This term improves, as expected, the audibility score, for all values $\lambda_2>0$. %we observe improvements in conventional captioning metrics, including BLEU~\cite{papineni2002bleu}, METEOR~\cite{banerjee2005meteor}, ROUGE-L~\cite{lin2004rouge}, and CIDEr~\cite{vedantam2015cider}. 
Overall, it is reassuring that our approach not only enhances audibility (our primary goal) but also positively impacts the overall quality of generated captions.
Sample results on the AudioCaps are presented in Fig.~\ref{fig:viz} and provide evidence of the tangible improvements attained through our approach. Of course, the video frames are given for visualization only and are not part of the input.

To further validate our method, we performed a user study. For this, we asked 30 users to evaluate captions for 10 random samples from the AudioCap datasets. After listening to the audio sample, the user was asked to select among the baseline ($\lambda_2=0$) and our method's caption ($\lambda_2=0.015$) according to two criteria: (i) which caption better describes the audio? and (ii) Which caption is more likely to arise from the audio alone, without additional information? 

The results indicate that in $86\% \pm 16$ of the cases, the users chose the caption produced by our method as the one that describes the audio, and in $85\% \pm 14$ of the cases they found that it better suited the contents of blind input.

\section{Limitations}
While our approach demonstrates promising results, it is important to acknowledge several limitations. Our methodology primarily focuses on generating captions that describe the audio content as a whole. It may not effectively identify or distinguish between multiple speakers or sources of sound within the same audio segment. This limitation is particularly relevant in scenarios involving conversations or multi-source audio. It is also evident that our solution does not address the task of speech transcription.

% No transcrption, Limited abilty to identify speakers and biased according to the web-extracted videos and caption in imagebond

%ImageBind, the multimodal alignment model guiding our approach, relies on the quality of caption data available in the web-extracted videos. In cases where captions in ImageBind are not accurate or informative, it can impact the overall performance of our audio captioning.

%Addressing these limitations and expanding the scope of our approach to handle scenarios with limited or no transcriptions and improved speaker identification are promising avenues for future research in the field of audio captioning.

\section{Conclusions}
Our pipeline offers a flexible and effective solution for zero-shot audio captioning, striking a balance between traditional captioning quality metrics and the crucial aspect of audibility. These results hold promise for a wide range of applications.

\bibliographystyle{IEEEbib}
\bibliography{caption}

\end{document}